\documentclass[
reprint,
dcolumn,
bm,
nofootinbib,
notitlepage]{revtex4-1}
\usepackage{graphicx}
\usepackage{color}
\usepackage[latin1]{inputenc}
\usepackage[T1]{fontenc}
\usepackage{amsmath}
\usepackage{amssymb}
\usepackage{braket}
\usepackage{mathrsfs}
\usepackage{hyperref}
\usepackage{ulem}
\usepackage{upgreek}
\usepackage{pstricks-add}
\usepackage{natbib} 
\usepackage{listings}
\usepackage{tikz}
\usepackage{tcolorbox}
\usetikzlibrary{mindmap}

\definecolor{lightblue}{rgb}{0.145,0.6666,1}

\begin{document}
\title{Cavity-induced Non-Adiabatic Dynamics and Spectroscopy of Molecular Rovibrational Polaritons studied by Multi-Mode Quantum Models}

\author{Eric W. Fischer}
\email{ericwfischer@posteo.de}
\affiliation{Theoretische Chemie, Institut f\"ur Chemie, Universit\"at Potsdam,
Karl-Liebknecht-Stra\ss{}e 24-25, D-14476 Potsdam-Golm, Germany}

\author{Peter Saalfrank}
\email{peter.saalfrank@uni-potsdam.de}
\affiliation{Theoretische Chemie, Institut f\"ur Chemie, Universit\"at Potsdam,
Karl-Liebknecht-Stra\ss{}e 24-25, D-14476 Potsdam-Golm, Germany}
\affiliation{Institut f\"ur Physik und Astronomie, Universit\"at Potsdam, Karl-Liebknecht-Stra\ss e 24-25, D-14476 Potsdam-Golm, Germany}

\date{\today}

\let\newpage\relax

\begin{abstract}
We study theoretically the quantum dynamics and spectroscopy of rovibrational polaritons formed in a model system composed of a single rovibrating diatomic molecule, which interacts with two degenerate, orthogonally polarized modes of an optical Fabry-P\'erot cavity. We employ an effective rovibrational Pauli-Fierz Hamiltonian in length gauge representation and identify three-state vibro-polaritonic conical intersections (VPCIs) between singly-excited vibro-polaritonic states in a two-dimensional angular coordinate branching space. The lower and upper vibrational polaritons are of mixed light-matter hybrid character, whereas the intermediate state is purely photonic in nature. The VPCIs provide effective population transfer channels between singly-excited vibrational polaritons, which manifest in rich interference patterns in rotational densities. Spectroscopically, three bright singly-excited states are identified, when an external infrared laser field couples to both a molecular and a cavity mode. The non-trivial VPCI topology manifests as pronounced multi-peak progression in the spectral region of the upper vibrational polariton, which is traced back to the emergence of rovibro-polaritonic light-matter hybrid states. Experimentally ubiquitous spontaneous emission from cavity modes induces a dissipative reduction of intensity and peak broadening, which mainly influences the purely photonic intermediate state peak as well as the rovibro-polaritonic progression.
\end{abstract}

\let\newpage\relax
\maketitle
\newpage

\section{Introduction}
\label{sec.introduction}
The interaction of infrared-active molecular vibrations with quantized electromagnetic field modes of optical Fabry-P\'erot-type cavities has been experimentally shown to have a peculiar impact on chemical ground state reactivity and molecular properties\cite{george2015,ebbesen2016,thomas2016,thomas2019,herrera2020}. The origin of this impact is attributed to the formation of vibrational light-matter hybrid states, known as vibrational polaritons\cite{ebbesen2016}. Linear\cite{thomas2016,thomas2019,chervy2018} and nonlinear\cite{xiang2018,zhang2019} infrared spectroscopic techniques have been the experimental methods of choice to characterize these vibrational light-matter hybrid states. Consequently, there has been a decent amount of theoretical effort to address the calculation of infrared spectra for vibro-polaritonic systems\cite{saurabh2016,ribeiro2018,flick2019,li2020,li2021,fischer2021,bonini2021}. There, one or more molecular vibrations interact with a single or multiple quantized cavity modes and the composite system is commonly described by an effective vibrational Pauli-Fierz Hamiltonian in the length-gauge representation of non-relativistic cavity quantum electrodynamics (cQED)\cite{flick2017a,flick2017b,schaefer2018,lihuo2021a}. In the latter, the light-matter interaction is mediated by the molecular ground state dipole moment and its orientation relative to the cavity mode polarization directions. Often, the dipole moment is assumed to be either exactly aligned with the polarization direction of a cavity mode or fixed at a certain orientation with respect to a single cavity mode 
polarization. 
The impact of molecular rotations on vibrational polaritons, which naturally influences the light-matter coupling due to the orientation dependent nature of the interaction, has been addressed only recently in a few works\cite{flick2018,sidler2020,szidarovszky2021,triana2021}. There, the orientation of the dipole moment was allowed to vary with respect to the polarization direction of a \textit{single} cavity mode and rotation related cavity-induced conical intersections have been studied in comparison to the role of classical laser fields\cite{triana2021}. However, the quantization of the cavity radiation field naturally leads to doubly degenerate optical modes with two orthogonal polarization directions\cite{thiru1998}, which both should be relevant for the light-matter interaction of rovibrating molecules. 

Here, we consider a rovibrating diatomic molecule in an optical cavity, which interacts with a pair of degenerate and orthogonally polarized cavity modes. Inspired by recent work of Vib\'ok and coworkers\cite{szidarovszky2021}, we treat the rovibro-polaritonic problem from a non-adiabatic perspective by energetically separating the rotational and vibro-polaritonic dynamics. As a result, we observe non-adiabatic effects that manifest as three-state vibro-polaritonic conical intersections (VPCIs) between vibro-polaritonic surfaces. We discuss the three-state VPCIs in detail from a topological, a dynamical and a spectroscopical perspective. Further, we examine the effect of experimentally ubiquitous dissipation on the infrared spectra of rovibrating molecules in an optical cavity due to spontaneous emission from excited states. We note the model character of our study, as gas phase experiments in optical cavities have not yet been performed due to issues in reaching the vibrational strong coupling regime. 

This paper is structured as follows: In Sec.\ref{sec.theory_model}, we introduce the effective rovibrational Pauli-Fierz Hamiltonian (\ref{subsec.rovi_pauli_fierz_hamiltonian}), our \textit{ab initio} molecular model (\ref{subsec.molecular_model}) and discuss the adiabatic separation of rotational and vibro-polaritonic degrees of freedom (\ref{subsec.adiabtic_separation}).
In Sec.\ref{sec.results_discussion}, we present and discuss our results. First, we examine the formation of three-state vibro-polaritonic conical intersections and their topological properties (\ref{subsec.rovib_lici}). Second, we discuss vibro-polaritonic population dynamics and the time-evolution of reduced vibro-polaritonic rotational densities (\ref{subsec.rovib_dynamics}). Third, we characterize the light-matter hybrid system by means of infrared spectra (\ref{subsec.rovib_spectra}) and discuss the spectroscopic manifestations of non-adiabatic features in comparison to purely vibro-polaritonic systems (\ref{subsubsec.rovib_ir_spectra}). Finally, we study the influence of dissipation on the infrared spectra due to spontaneous emission effects(\ref{subsubsec.lossy_cavity_rovib_spectra}). Sec.\ref{sec.summary} summarizes our work and connects our findings to other branches of molecular polaritonics.
\section{Theory and Model}
\label{sec.theory_model}
\subsection{Rovibrational Pauli-Fierz Hamiltonian}
\label{subsec.rovi_pauli_fierz_hamiltonian}
\begin{figure}[hbt]
\includegraphics[scale=1.0]{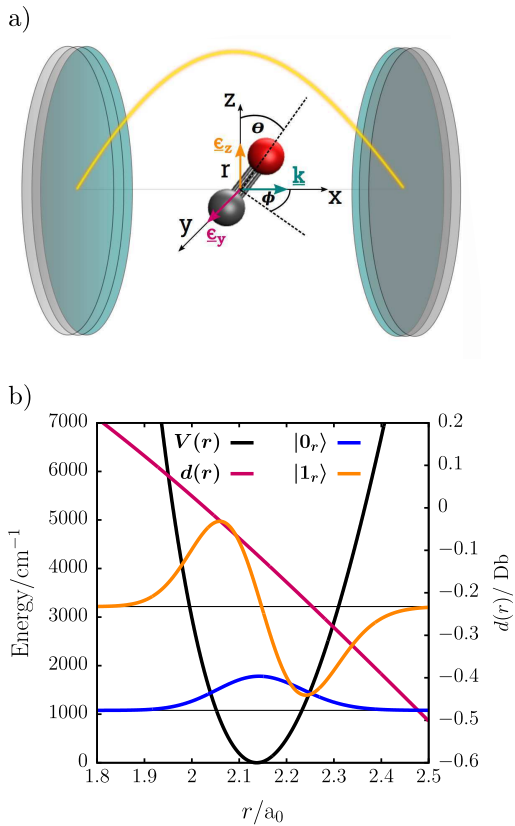}
\renewcommand{\baselinestretch}{1.}
\caption{(a) Schematic sketch of diatomic molecule with vibrational coordinate, $r$, and angular coordinates $\theta,\phi$ in optical two-mode cavity with polarization vectors $\underline{\epsilon}_z,\underline{\epsilon}_y$, wavevector, $\underline{k}$, and molecular space-fixed frame with axes, $x,y,z$. A $z$-polarized cavity mode is indicated in yellow. (b) Anharmonic potential, $V(r)$, and dipole function, $d(r)$, of CO stretching mode (interpolated CCSD(T)/aug-cc-pV5Z results) with vibrational ground state, $\ket{0_r}$, and first excited state, $\ket{1_r}$. $d(r)$ is given in units of Debye (Db).}
\label{fig.rovibcavity_diatomic_system}
\end{figure}

We consider a diatomic molecule in a two-mode optical cavity (\textit{cf.} Fig.\ref{fig.rovibcavity_diatomic_system}(a)) described by an effective length-gauge, rovibrational Pauli-Fierz Hamiltonian in dipole and cavity Born-Oppenheimer type approximation\cite{fischer2021,flick2017a,flick2017b,schaefer2018}, as given by
\begin{equation}
\hat{H}
=
\hat{H}_S
+
\hat{H}_C
+
\hat{H}_{SC}
+
\hat{H}_{DSE}
\quad.
\label{eq.rovib_pauli_fierz}
\end{equation}
In the specific example chosen in this work, $\hat{H}_S$ is the Hamiltonian of a single rovibrating CO molecule 
\begin{equation}
\hat{H}_S
=
\dfrac{\hat{j}^2}{2 I}\,
\underbrace{-
\dfrac{\hbar^2}{2\mu}
\dfrac{\partial^2}{\partial r^2}
+
V(r)}_{=\hat{H}_\mathrm{vib}}
\quad,
\label{eq.rovib_hamiltonian}
\end{equation}
with moment of inertia, $I=\mu r^2$, where $r$ is the CO-stretching coordinate and $\mu$ the reduced mass, and angular momentum operator 
\begin{equation}
\hat{j}^2
=
-
\hbar^2
\dfrac{1}{\sin\theta}
\dfrac{\partial}{\partial \theta}\,
\sin\theta
\dfrac{\partial}{\partial \theta}
+
\dfrac{1}{\sin^2\theta}\dfrac{\partial^2}{\partial \phi^2}
\quad,
\end{equation} 
with polar angle, $\theta\in[0,\pi]$, and azimuthal angle, $\phi\in[0,2\pi)$, respectively. The vibrational Hamiltonian, $\hat{H}_\mathrm{vib}$, is determined by the molecular Born-Oppenheimer ground state potential energy surface, $V(r)$. 

Further, we consider a Fabry-P\'erot type cavity with two energetically degenerate and orthogonally polarized, transverse cavity modes 
\begin{equation}
\hat{H}_C
=
\sum_{\lambda=z,y}
\hbar\omega_c
\left(
\hat{a}^\dagger_\lambda
\hat{a}_\lambda
+
\dfrac{1}{2}
\right) \quad,
\label{eq.closed_cavity}
\end{equation}
with polarization index $\lambda$, harmonic cavity frequency $\omega_c$ and photon creation/annihilation operators $\hat{a}^\dagger_\lambda/\hat{a}_\lambda$, respectively. Here, we chose the cavity modes to be polarized along the $z$- and $y$-axis of the molecular space-fixed frame (perpendicular to the wave vector, $\underline{k}$), as shown in Fig.\ref{fig.rovibcavity_diatomic_system}(a). Further, the orientation dependent light-matter interaction is mediated by the molecular dipole moment
\begin{equation}
\underline{d}(r,\theta,\phi)
=
d(r)
\begin{pmatrix}
\sin\theta\cos\phi
\vspace{0.1cm}\\
\sin\theta\sin\phi
\vspace{.1cm}\\

\cos\theta
\end{pmatrix} 
\quad,
\end{equation}
where $d(r)$ is the molecular dipole moment along the CO-bonding axis. The light-matter interaction is given by the projection of $\underline{d}(r,\theta,\phi)$ on the cavity mode polarization directions, $\left(\underline{\epsilon}_\lambda\cdot\underline{d}(r,\theta,\phi)\right)$, with $\lambda=z,y$. Accordingly, the bare cavity-molecule interaction Hamiltonian reads
\begin{align}
\hat{H}_{SC}
&=
g
\sum_{\lambda=z,y}
\biggl(\underline{\epsilon}_\lambda\cdot\underline{d}(r,\theta,\phi)\biggr)
\left(
\hat{a}^\dagger_\lambda
+
\hat{a}_\lambda
\right)
\quad,
\label{eq.bare_light_matter}
\vspace{0.2cm}
\\
&=
g\,d(r)
\biggl(
\left(
\hat{a}^\dagger_z
+
\hat{a}_z
\right)
\cos\theta
+
\left(
\hat{a}^\dagger_y
+
\hat{a}_y
\right)
\sin\theta\sin\phi
\biggr) 
\quad,
\nonumber
\end{align}
and the dipole self-energy, which has been thoroughly studied in Refs.\cite{rokaj2018} and\cite{schaefer2020}, is given by
\begin{align}
\hat{H}_{DSE}
&=
\dfrac{g^2}{\hbar\omega_c}
\sum_{\lambda=z,y}
\biggl(\underline{\epsilon}_\lambda\cdot\underline{d}(r,\theta,\phi)\biggr)^2
\quad,
\label{eq.dipole_self_energy}
\vspace{0.2cm}
\\
&=
\dfrac{g^2}{\hbar\omega_c}
d^2(r)
\biggl(
\cos^2\theta
+
\sin^2\theta\sin^2\phi
\biggr)
\quad.
\nonumber
\end{align}
Note, there are two independent contributions to the DSE term related to independent z- and y-polarized cavity modes without any cross terms coupling modes along \textit{different} polarization directions.

The light-matter interaction strength, $g$, in Eqs.\eqref{eq.bare_light_matter} and \eqref{eq.dipole_self_energy}, carrying the dimension of an electric field strength (V/m), is related to a dimensionless parameter $\eta$ via\cite{kockum2019}
\begin{equation}
g
=
\dfrac{\hbar\omega_c}{\vert d_{fi}\vert}\,\eta
\quad .
\end{equation}
Here, $d_{fi}$ is a  vibrational transition dipole matrix element between selected initial and final molecular states (see below), to which we choose the cavity mode frequency $\omega_c$ to be resonant. Generally, the coupling parameter $g$ (or $\eta$) is a function of the cavity volume, the dielectric constant inside the cavity and the number of molecules in the cavity.\cite{fischer2021} Here we treat $g$ (and $\eta$) as an adjustable parameter. Further, $\eta$ specifies the light-matter interaction regime: Vibrational strong coupling (VSC) for $0.0<\eta<0.1$ and vibrational ultrastrong coupling (VUSC) for $\eta\geq0.1$, respectively.\cite{kockum2019}

In what follows, by assuming that the center of mass of the rovibrating CO is fixed in the cavity, the full Hamiltonian is five-dimensional (three molecular coordinates, $r$, $\theta$ and $\phi$, and two cavity modes). Below we shall work in a mixed basis to represent $\hat{H}$, namely a grid basis for the angular coordinates and a state-representation for the vibro-polaritonic modes as characterized by quantum numbers $v_r$ (for the CO-vibration) and ($n_z$, $n_y$) for the cavity modes. Further, we emphasize the CO molecule's free rotation is described by two angular coordinates $\theta,\phi$, \textit{i.e.}, the CO molecule is allowed to take arbitrary orientations with respect to the cavity polarization plane spanned by vectors, $\underline{\epsilon}_z$ and $\underline{\epsilon}_y$. Moreover, we do not take into account the molecule's center of mass motion as cavity boundary effects are assumed to play no role under the long-wavelength approximation and the neutral CO's center of mass motion does not directly couple to the cavity modes, which would require a net molecular charge\cite{sidler2020}.

The dynamics as generated by the rovibrational Pauli-Fierz Hamiltonian in Eq.\eqref{eq.rovib_pauli_fierz}, is governed by a time-dependent Schr\"odinger equation
\begin{equation}
\mathrm{i}\hbar\,
\dfrac{\partial}{\partial t}
\Psi(v_r,n_z,n_y,\theta,\phi,t)
=
\hat{H}\,
\Psi(v_r,n_z,n_y,\theta,\phi,t)
\quad,
\label{eq.time_dependent_schroedinger_equation}
\end{equation}
with a five-dimensional rovibro-polaritonic wave packet, $\Psi(v_r,n_z,n_y,\theta,\phi,t)$. We will discuss two different initial states to solve Eq.(\ref{eq.time_dependent_schroedinger_equation}), as specified below, given by an isolated rovibrational excitation of the molecule and a light-matter superposition state, respectively.

\subsection{Molecular Model}
\label{subsec.molecular_model}
We study a single carbon monoxide molecule with reduced mass, $\mu=m_\mathrm{C}m_\mathrm{O}/m_\mathrm{CO}=12506\,m_e$ (electron mass $m_e$). The molecular \textit{ab initio} PES, $V(r)$, and molecular dipole function, $d(r)$, were calculated as function of the CO-bond-length $r$ (\textit{cf.} Fig.\ref{fig.rovibcavity_diatomic_system}(b)) on the CCSD(T)/aug-cc-pV5Z level of theory via the software package Gaussian16\cite{g16}. We find a ground state equilibrium bond length of $r_e=2.145\,a_0$ and a rotational constant of $B=\frac{\hbar}{4\pi c\mu r^2_e}=1.91\,\mathrm{cm}^{-1}$, which is in close agreement with the experimental value of $B_\mathrm{exp}=1.92\,\mathrm{cm}^{-1}$\cite{rank1965}.

Further, we numerically obtained the two lowest vibrational eigenvalues/eigenstates (\textit{cf.} Fig.\ref{fig.rovibcavity_diatomic_system}(b)) of the vibrational Hamiltonian, $\hat{H}_\mathrm{vib}$, in Eq.\eqref{eq.rovib_hamiltonian} based on a Colbert-Miller discrete variable representation\cite{colbert1992} with $N_r=1501$ grid points. The fundamental anharmonic transition frequency, $\hbar\omega_{10}=2137\,\mathrm{cm}^{-1}$, between the two lowest lying vibrational eigenstates of the CO stretching mode compares well to an experimental value of $2143\,\mathrm{cm}^{-1}$\cite{camilde1996}. 
The corresponding vibrational transition dipole moment takes $d_{10}=0.066$ ea$_0$, which is in agreement with Ref.\cite{huajun2015}. The molecular dipole function $d(r)$ changes roughly linearly in the $r$-range shown in Fig.\ref{fig.rovibcavity_diatomic_system}(b) and takes an absolute equilibrium value of $\vert d(r_e)\vert=0.12\,\mathrm{Db}$, which is in close agreement with literature\cite{scuseria1991}. In the following, we set the cavity mode frequency $\hbar\omega_c=\hbar\omega_{10}$ and the transition dipole moment $d_{fi}=d_{10}$.
\subsection{Diabatic Vibro-Polaritonic Basis}
\label{subsec.adiabtic_separation}
The rotational constant, $B$, and the fundamental vibrational transition energy, $\hbar\omega_{10}$, set two different excitation energy scales, which allow for adiabatic separation of vibrational (``fast'') and rotational (``slow'') degrees of freedom\cite{szidarovszky2021}. Accordingly, we consider a restricted basis of zero-order ``vibro-polaritonic'' states, $\ket{v_r,n_z,n_y}$, which constitute eigenstates to $\hat{H}_\mathrm{vib}+\hat{H}_C$, and is given by
\begin{equation}
\begin{cases}
X_0: & \ket{0_r,0_z,0_y}
\vspace{0.2cm}
\\
X_1: & \ket{1_r,0_z,0_y},\ket{0_r,1_z,0_y},\ket{0_r,0_z,1_y}
\end{cases}
\quad,
\label{eq.vibrational_zero_order_basis}
\end{equation}
with one-dimensional ground state manifold, $X_0$, and three-dimensional singly-excited state manifold, $X_1$. The $X_1$-manifold allows for the description of the lowest lying excited vibro-polaritonic states in the VSC regime. Higher-lying excited states are neglected here. 

In the following, we denote the four basis states in Eq.\eqref{eq.vibrational_zero_order_basis} generically as $\ket{D_k}$ and expand a rovibro-polaritonic wave packet as
\begin{equation}
\Psi(v_r,n_z,n_y,\theta,\phi,t)
=
\sum_k
\varphi_k(\theta,\phi,t)
\ket{D_k}
\quad,
\label{eq.rovibpol_wavepacket}
\end{equation}
with time-dependent, rotational wave packets, $\varphi_k(\theta,\phi,t)$. Further, the matrix representation of the rovibrational Pauli-Fierz Hamiltonian in the zero-order basis, which constitutes a matrix operator in $(\theta,\phi)$-space, is given by
\begin{equation}
\underline{\underline{H}}
=
\underline{\underline{T}}(\theta,\phi)
+
\underline{\underline{V}}(\theta,\phi)
\quad ,
\label{eq.matrix_rovib_pauli_fierz_hamiltonian}
\end{equation}
with ($4\times4$)-blocked potential energy, $\underline{\underline{V}}(\theta,\phi)$, and rotational kinetic energy operator matrices, $\underline{\underline{T}}(\theta,\phi)$, respectively. The latter is given by 
\begin{widetext}
\begin{align}
\underline{\underline{{T}}}(\theta,\phi)
=
\frac{\hat{j}^2}{2 \mu}  \begin{pmatrix}
 \langle 0_r | \frac{1}{r^2} | 0_r \rangle  & \langle 0_r | \frac{1}{r^2} | 1_r \rangle & 0 & 0 \\
 \langle 1_r | \frac{1}{r^2} | 0_r \rangle  & \langle 1_r | \frac{1}{r^2} | 1_r \rangle & 0 & 0 \\
 0 & 0 & \langle 0_r | \frac{1}{r^2} | 0_r \rangle & 0 \\
 0 & 0 & 0 & \langle 0_r | \frac{1}{r^2} | 0_r \rangle &  
\\
\end{pmatrix}
\label{eq.kinetic_matrix}
\end{align}
\end{widetext}
where $\langle \dots \rangle$ indicates integration with respect to the $r$-coordinate. The zero-point energy shifted potential energy matrix is given by
\begin{widetext}
\begin{align}
\underline{\underline{V}}(\theta,\phi)
=
\begin{pmatrix}
\dfrac{g^2}{\hbar\omega_c}f_{\theta\phi}\braket{d^2}_{00} & \dfrac{g^2}{\hbar\omega_c}f_{\theta\phi}\braket{d^2}_{10} & g\,d_{00}\cos\theta  & g\,d_{00}\sin\theta\sin\phi\\
\dfrac{g^2}{\hbar\omega_c}f_{\theta\phi}\braket{d^2}_{10} & \hbar\omega_{10}+\dfrac{g^2}{\hbar\omega_c}f_{\theta\phi}\braket{d^2}_{11} & g\,d_{10}\cos\theta & g\,d_{10}\sin\theta\sin\phi\\
g\,d_{00}\cos\theta          & g\,d_{10}\cos\theta & \hbar\omega_c+\dfrac{g^2}{\hbar\omega_c}f_{\theta\phi}\braket{d^2}_{00} & 0\\
g\,d_{00}\sin\theta\sin\phi         & g\,d_{10}\sin\theta\sin\phi & 0 & \hbar\omega_c+\dfrac{g^2}{\hbar\omega_c}f_{\theta\phi}\braket{d^2}_{00}\\
\end{pmatrix}
\quad,
\label{eq.potential_matrix}
\end{align}
\end{widetext}
with dipole self-energy matrix elements, $\braket{d^2}_{v_r v^\prime_r}=\braket{v_r\vert d^2(r)\vert v^\prime_r}$, and
\begin{equation}
f_{\theta\phi}
=
\cos^2\theta+\sin^2\theta\sin^2\phi
\quad.
\end{equation} 
In $\underline{\underline{V}}(\theta,\phi)$, the first diagonal entry corresponds to the zero-order ground state and the diagonal elements of the lower $3\times3$-block resemble states forming the $X_1$-manifold in the order given by Eq.\eqref{eq.vibrational_zero_order_basis}. (The same 
 assignment can be made for the kinetic-energy matrix elements in Eq.(\ref{eq.kinetic_matrix}).)

The $X_0$- and $X_1$-manifolds are energetically well separated by $\hbar\omega_{10}=\hbar\omega_c$ and subject to a weak anharmonic coupling ($g^2\braket{d^2}_{10}$) induced by the dipole self-energy and comparatively strong light-matter coupling ($g\,d_{00}$). Further, the DSE constitutes a potential in angular coordinates on the diagonal ($g^2\,\braket{d^2}_{v_rv_r}f_{\theta\phi}$) and the singly-excited molecular state couples to the two singly-excited cavity mode states ($ g\,d_{10}$). 

There is only a weak kinetic coupling in the kinetic-energy matrix, by the term $\frac{\hat{j}^2}{2\mu} \braket{0_r\vert\frac{1}{r^2}\vert 1_r}$. Hence, the Pauli-Fierz Hamiltonian takes a diabatic-like representation in the zero-order basis Eq.\eqref{eq.vibrational_zero_order_basis}. Following arguments given in Ref.\cite{szidarovszky2021}, we interpret the eigenstates of $\underline{\underline{V}}(\theta,\phi)$ as vibro-polaritonic states, which are given by the vibro-polaritonic ground state $\ket{G}$, the lower $\ket{L_1}$, the middle $\ket{M_1}$ and the upper $\ket{U_1}$ vibro-polaritonic states, respectively. Note, the latter take the role of adiabatic states here. Further, the corresponding eigenvalues $\varepsilon_{G}(\theta,\phi),\varepsilon_{L_1}(\theta,\phi),\varepsilon_{M_1}(\theta,\phi)$ and $\varepsilon_{U_1}(\theta,\phi)$, are functions of the angular coordinates, $\theta$ and $\phi$, and constitute vibro-polaritonic potential energy surfaces (PES) for the rotational dynamics of the molecule.

\section{Results and Discussion}
\label{sec.results_discussion}
\subsection{Vibro-Polaritonic Conical Intersections}
\label{subsec.rovib_lici}
\begin{figure*}[hbt]
\includegraphics[scale=1.0]{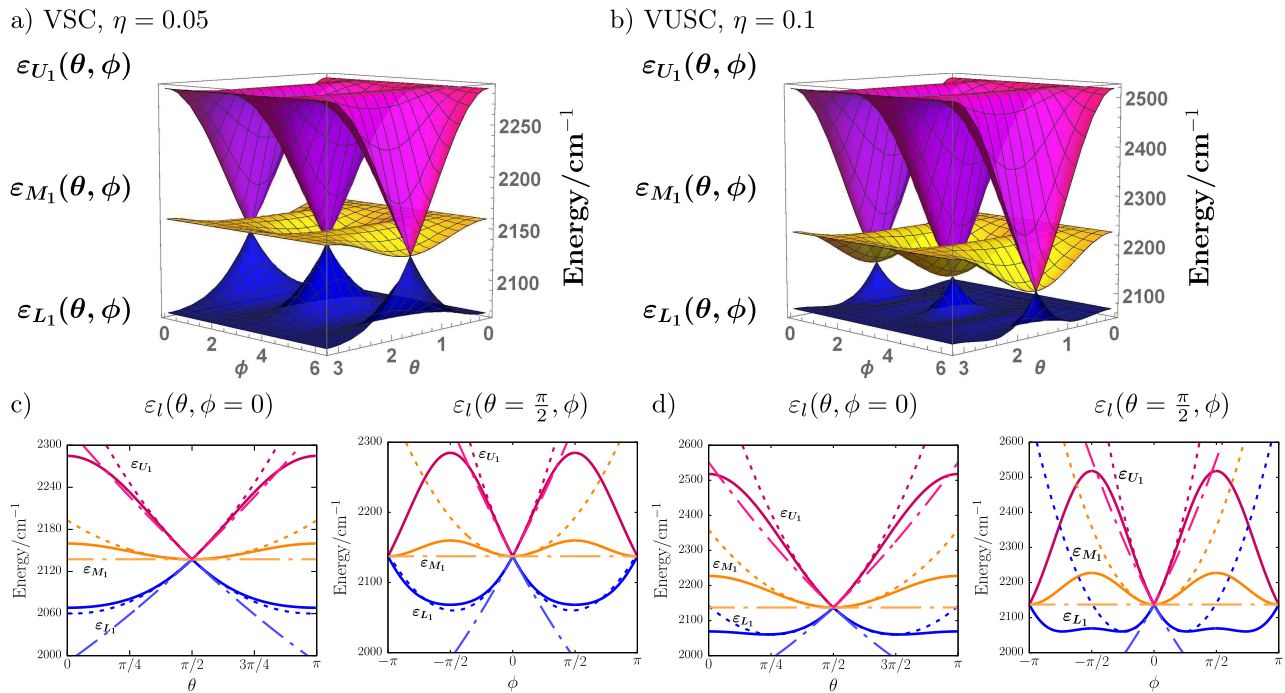}
\caption{Top: Three-state vibro-polaritonic conical intersections (VPCIs) between lower vibro-polaritonic, $\varepsilon_{L_1}(\theta,\phi)$, middle vibro-polaritonic, $\varepsilon_{M_1}(\theta,\phi)$, and upper vibro-polaritonic surfaces, $\varepsilon_{U_1}(\theta,\phi)$, for (a) VSC regime with $\eta=0.05$ and (b) onset of VUSC regime with $\eta=0.1$. Bottom: One dimensional cuts through vibro-polaritonic surfaces (bold), linear (dotted-dashed) and quadratic approximations (dotted) with $\varepsilon_l(\theta,\phi=0)$ and $\varepsilon_l(\theta=\frac{\pi}{2},\phi)$ for (c) $\eta=0.05$ and (d) $\eta=0.1$ with $l=L_1,M_1,U_1$.}
\label{fig.lici_free_vsc_fig}
\end{figure*}

We first discuss cavity-mode induced non-adiabatic effects in the manifold of singly-excited vibro-polaritonic states. Under vibrational strong coupling, we identify three-state vibro-polaritonic conical intersections (VPCIs) between vibro-polaritonic surfaces $\varepsilon_{L_1}(\theta,\phi)$, $\varepsilon_{M_1}(\theta,\phi)$ and $\varepsilon_{U_1}(\theta,\phi)$ under the conditions
\begin{equation}
\hbar\omega_{10}
=
\hbar\omega_c,
\hspace{0.3cm}
\theta
=
\dfrac{\pi}{2},
\hspace{0.3cm}
\phi
=
0,\pi
\quad.
\label{eq.lici_conditions}
\end{equation}
The three-state VPCIs, as shown in Figs.\ref{fig.lici_free_vsc_fig}(a) and (b) for $\eta=0.05$ and $\eta=0.1$, respectively, are located in a two-dimensional angular coordinate branching space. Two distinct intersections are identified with intersection coordinates $(\theta,\phi)=(\frac{\pi}{2},0)$ and $(\frac{\pi}{2},\pi)$, due to the periodicity of the azimuthal angle, $\phi$.

The three-state VPCIs exhibit a characteristic double-cone topology formed by the $L_1$- and $U_1$-surfaces. They are triply degenerate at the intersection point due to an additional crossing with the middle vibro-polaritonic surface, $\varepsilon_{M_1}(\theta,\phi)$, which exhibits a local minimum here. In the VSC regime ($\eta=0.05$), the splitting of the $L_1$- and $U_1$-surfaces is slightly asymmetric with respect to the $M_1$-surface as can be seen in Figs.\ref{fig.lici_free_vsc_fig}(a) and (c). At the onset of the VUSC regime ($\eta=0.1$), the $L_1$/$U_1$-splitting turns out to be strongly asymmetric (\textit{cf.} Figs.\ref{fig.lici_free_vsc_fig}(b) and (d)), with a dominant inverse cone in $\varepsilon_{U_1}(\theta,\phi)$. The $M_1$-surface forms more pronounced minima at the intersection points for stronger light-matter interaction and the $L_1$-surface is subject to a mild ``Mexican-hat''-type topology close to the intersection coordinate.

To reveal detailed characteristics of the VPCI close to the intersection coordinate, we expand $\underline{\underline{V}}(\theta,\phi)$ in $\theta,\phi$ around the intersection point at $(\theta,\phi)=(\frac{\pi}{2},0)$ (\textit{cf.} Appendix B for details) in analogy to vibronic problems. In Figs.\ref{fig.lici_free_vsc_fig}(c) and (d), we show cuts through the VPCIs along $\theta$ and $\phi$, with exact surfaces (bold) besides linear (dotted-dashed) and quadratic (dotted) approximations. We observe three characteristic features of the VPCI: (i) the degeneracy is lifted linearly along both angular coordinates in $\varepsilon_{L_1}(\theta,\phi)$ and $\varepsilon_{U_1}(\theta,\phi)$ in analogy to linear vibronic coupling theory\cite{koeppel1984}, (ii) $\varepsilon_{M_1}(\theta,\phi)$ exhibits a harmonic character at the intersection point, which results from the potential character of the dipole self-energy and (iii) the mild local ``Mexican-hat''-type topology of $\varepsilon_{L_1}(\theta,\phi)$ at $\eta=0.1$ also stems from the DSE.

We close by noting, that the topological motive of a three-state VPCI is similar to \textit{accidental} three-state conical intersections in molecular vibronic coupling theory.\cite{matsika2003,matsika2005,coe2005,matsika2011}

\subsection{Rovibro-Polaritonic Dynamics}
\label{subsec.rovib_dynamics}
We now turn to the dynamics of different zero-order rovibro-polaritonic wave packets, which mimic a light-matter hybrid system initially excited by an external classical, $z$-polarized laser field. The additional action of a laser could be included in the Pauli-Fierz Hamiltonian by adding a term, $- \underline{\mu}\cdot\underline{E}(t)$ (with classical laser field $\underline{E}(t)$ and where $\underline{\mu}$ is a ``dipole'' function depending on molecular and cavity degrees of freedom\cite{flick2019}), but here we simply assume that such an excitation already took place and appropriate initial rovibrational(-polaritonic) states have been prepared. In order to represent different laser-polariton coupling scenarios, or, implicitly, molecule-cavity dipole functions, we consider two different initial states, namely
\begin{align}
\Psi^{(r)}_0
&=
\ket{1_r, 0_z, 0_y}\,Y^1_0(\theta,\phi)
\quad \mathrm{and}
\label{eq.init_molxite}
\vspace{0.2cm}
\\
\Psi^{(r,z)}_0
&=
\frac{1}{\sqrt{2}}\biggl(\ket{1_r, 0_z,0_y}+\ket{0_r, 1_z,0_y}\biggr)\,Y^1_0(\theta,\phi)
\,,
\label{eq.init_supxite}
\end{align}
with first excited rotational state $Y^1_0(\theta,\phi)$, where we adopt the notation, $Y^j_{m_j}(\theta,\phi)$, for spherical harmonics. Here, $\Psi^{(r)}_0$ resembles a purely molecular rovibrational excitation and $\Psi^{(r,z)}_0$ constitutes a superposition of a singly-excited molecular and a singly-excited cavity mode state, respectively. The first case resembles a scenario, where an external laser field had coupled exclusively to the molecule, whereas the latter considers additionally a coupling of the laser-field to the $z$-polarized cavity mode.\cite{rovibXite} 

Using these initital states, we discuss the time-evolution of the light-matter hybrid system in terms of zero-order (diabatic) populations 
\begin{align}
P^{(k)}_\mathrm{dia}(t)
&=
\int^{2\pi}_0\int^\pi_0
\vert\varphi_k(\theta,\phi,t)\vert^2\,
\sin\theta\,
\mathrm{d}\theta\,
\mathrm{d}\phi
\quad,
\label{eq.diabatic_populations}
\end{align}
with $\varphi_k$ given in Eq.(\ref{eq.rovibpol_wavepacket}), and their vibro-polaritonic (adiabatic) counterparts, $P^{(l)}_\mathrm{ad}(t)$, with $l=G,L_1,M_1,U_1$. The latter arise from adiabatic wave packets, $\varphi^\mathrm{ad}_l(\theta,\phi,t)$, by using an expansion analogous to Eq.(\ref{eq.rovibpol_wavepacket} but with adiabatic basis states $\ket{G}$, $\ket{L_1}$, $\ket{M_1}$ and $\ket{U_1}$. Further, we study the adiabatic rotational dynamics by means of reduced rotational densities
\begin{equation}
\rho^\mathrm{ad}_l(\theta,t) 
=
\int^{2\pi}_0
\vert\varphi^\mathrm{ad}_l(\theta,\phi,t)\vert^2\, 
\mathrm{d}\phi
\quad ,
\label{eq.adiabatic_theta_density}
\end{equation}
which resemble the reduced dynamics of rotational wave packets, $\varphi^\mathrm{ad}_l(\theta,\phi,t)$, on the vibro-polaritonic surfaces as depicted in Fig.\ref{fig.lici_free_vsc_fig}. We note, that dynamics along the $\theta$-coordinate turns out to be particularly illustrative compared to the $\phi$-coordinate. 

We solve the TDSE \eqref{eq.time_dependent_schroedinger_equation} with the multiconfigurational time-dependent Hartree (MCTDH) method in its multi-set formulation\cite{meyer2012} as implemented in the Heidelberg MCTDH package \cite{mctdh2019}. Numerical details on the MCTDH method as well as on evaluation of above mentioned populations and reduced densities are provided in Appendix A.

\subsubsection{Vibro-Polaritonic Population Dynamics}
\label{subsubsec.vibpol_dynamics}
\begin{figure}[hbt]
\includegraphics[scale=1.0]{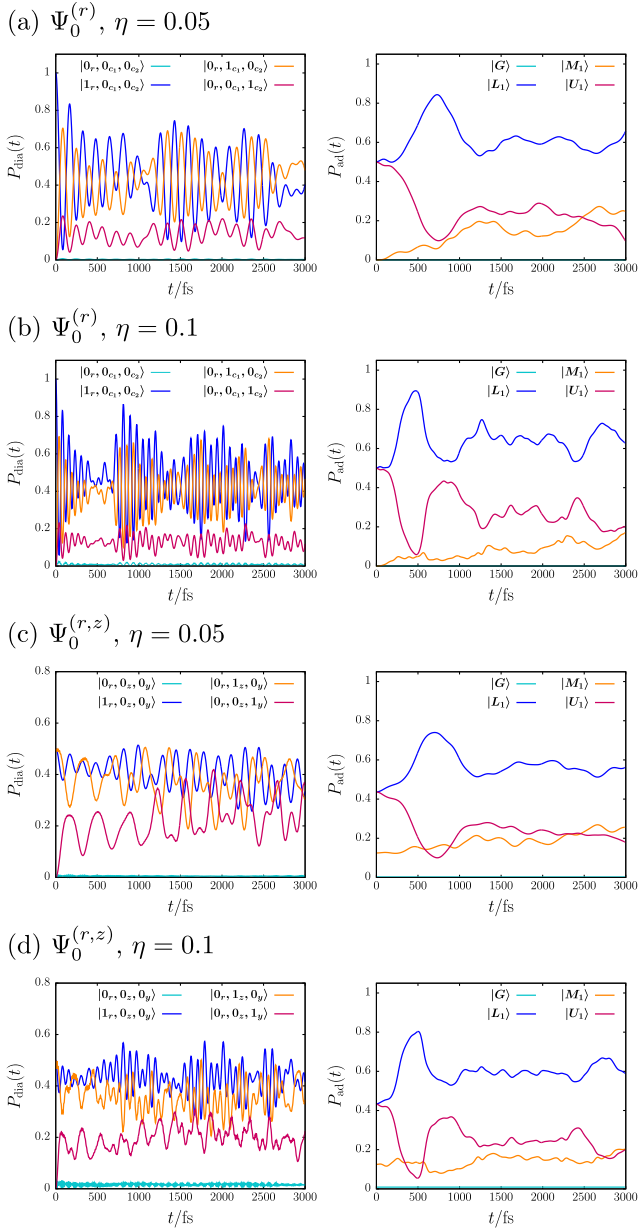}
\renewcommand{\baselinestretch}{1.}
\caption{Population dynamics for different initial states under VSC. Zero-order (diabatic), $P_\mathrm{dia}(t)$, (left) and vibro-polaritonic (adiabatic), $P_\mathrm{ad}(t)$, (right) population dynamics for rovibrational singly-excited initial state, $\Psi^{(r)}_0$, for (a) $\eta=0.05$ and (b) $\eta=0.1$. Zero-order (left column) and vibro-polaritonic (right column) population dynamics for singly-excited superposition state, $\Psi^{(r,z)}_0$, for (c) $\eta=0.05$ and (d) $\eta=0.1$.}
\label{fig.rovib_polariton_dynamics}
\end{figure}

Before we discuss the time-evolution of populations, we examine zero-order state contributions to stationary vibro-polaritonic states at $t=0$. For the rovibrationally excited initial state, $\Psi^{(r)}_0$, we find only $\ket{L_1}$ and $\ket{U_1}$ equally populated for both light-matter interaction scenarios with $\eta=0.05$ and $\eta=0.1$ (\textit{cf.} Figs.\ref{fig.rovib_polariton_dynamics}(a) and (b)). 
In particular, we find no population in the middle vibro-polaritonic state $\ket{M_1}$, \textit{i.e.}, no molecular vibrationally excited state is contributing here, which identifies $\ket{M_1}$ as purely photonic in character. Further, for the superposition state, $\Psi^{(r,z)}_0$, we observe contributions from all three vibro-polaritonic states, with equally populated $\ket{L_1}$ and $\ket{U_1}$ constituting the dominant contribution opposed to $\ket{M_1}$ (\textit{cf.} Figs.\ref{fig.rovib_polariton_dynamics}(c) and (d)). Accordingly, $\ket{L_1}$ and $\ket{U_1}$ are identified as ``true'' light-matter hybrid states containing both molecular and photonic contributions. 

Turning to the dynamics, we consider a total propagation time of $t_f=3000\,\mathrm{fs}$ for zero-order (``diabatic'') populations, $P_\mathrm{dia}(t)$, and their vibro-polaritonic (``adiabatic'') counterparts, $P_\mathrm{ad}(t)$. We find the dynamics up to $t_f$ to be mainly dominated by states spanning the $X_1$-manifold as shown in the left and right columns of Fig.\ref{fig.rovib_polariton_dynamics}, which results from the relatively large energetic separation of $X_0$- and $X_1$-manifolds.

For the zero-order (``diabatic'') dynamics initiated by $\Psi^{(r)}_0$ and depicted in Figs.\ref{fig.rovib_polariton_dynamics}(a) and (b) (left column), we observe a fast initial population transfer to both singly-excited cavity mode states followed by  coherent exchange dynamics. The latter are dominated by the molecular and the $z$-polarized excited states and subject to quantum beats with a period of roughly $1000\,\mathrm{fs}$. For an increased light-matter interaction strength ($\eta=0.1$), the quantum beat period shortens to roughly $600\,\mathrm{fs}$ and the frequency of coherent zero-order population transfer is strongly enhanced. In contrast, vibro-polaritonic (``adiabatic'') dynamics (\textit{cf}. Figs.\ref{fig.rovib_polariton_dynamics}(a) and (b), right column) are dominated by slow population transfer from $\ket{U_1}$ to $\ket{L_1}$, accompanied by a gradual population of $\ket{M_1}$. For $\eta=0.05$ we observe a characteristic maximum in the $\ket{L_1}$-population at around $700\,\mathrm{fs}$, which is shifted to roughly $500\,\mathrm{fs}$ for $\eta=0.1$. Later times are characterized by several decaying recurrences in $\ket{U_1}$.

For the dynamics initiated by $\Psi^{(r,z)}_0$, we find a significantly slower zero-order population transfer between the molecular and the $z$-polarized excited mode states for both coupling scenarios (\textit{cf.} Figs.\ref{fig.rovib_polariton_dynamics}(c) and (d), left column). From the vibro-polaritonic perspective, as depicted in the right column of Figs.\ref{fig.rovib_polariton_dynamics}(c) and (d), $\ket{U_1}$ is initially again depopulated in favor of $\ket{L_1}$, accompanied by a slight population increase in $\ket{M_1}$. The first maximum in $\ket{L_1}$ appears at same times as observed above and for longer times, the population dynamics is less structured here. We  note, for both $\Psi^{(r)}_0$ and $\Psi^{(r,z)}_0$ the zero-order ground state is weakly contributing at $\eta=0.1$ and exhibits a strongly oscillatory dynamics, which is especially pronounced for $\Psi^{(r,z)}_0$. 

Finally, from this population-based perspective, we conclude that (i) only $\ket{L_1}$ and $\ket{U_1}$ are light-matter hybrid states opposed to the purely photonic $\ket{M_1}$ state and (ii) the inclusion of rotational degrees of freedom provides a cavity-induced transfer channel between vibro-polaritonic excited states. Notably, the latter is absent in purely vibrational problems, where an additional bath is required to mediate population transfer between vibro-polaritonic states.

\subsubsection{Reduced Rotational Dynamics}
\label{subsubsec.rotational_dynamics}
We now turn to rotational dynamics of the CO molecule based on vibro-polaritonic (``adiabatic'') reduced rotational densities, $\rho^\mathrm{ad}_l(\theta,t)$ (\textit{cf}. Eq.\eqref{eq.adiabatic_theta_density}), for different vibro-polaritonic surfaces. In order to distinguish reduced densities for different initial states, we introduce the notation $\rho^{(r)}_l(\theta,t)$ for $\Psi^{(r)}_0$ and $\rho^{(r,z)}_l(\theta,t)$ for $\Psi^{(r,z)}_0$, respectively.

\begin{figure}[hbt]
\includegraphics[scale=1.0]{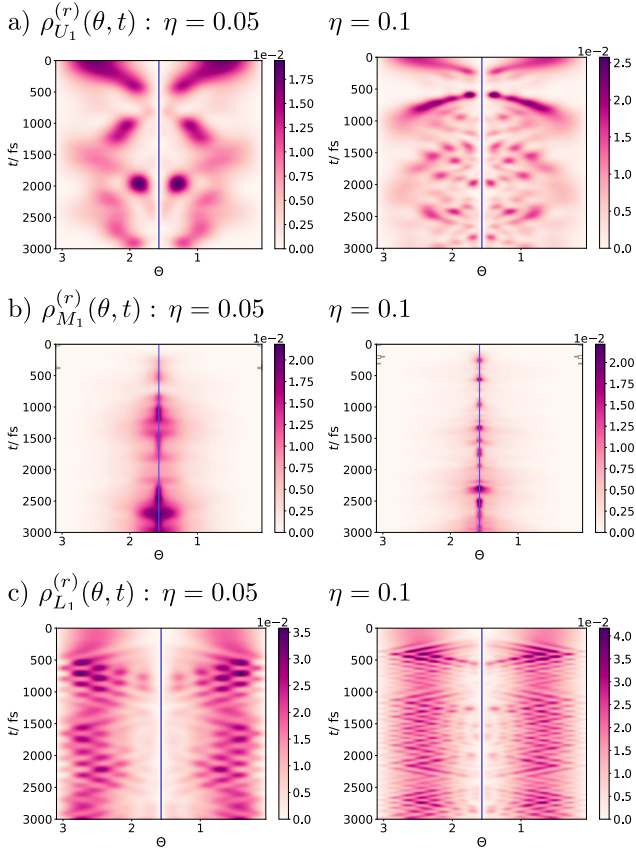}
\renewcommand{\baselinestretch}{1.}
\caption{Time-evolution of adiabatic reduced rotational densities, $\rho^{(r)}_l(\theta,t)$, for rovibrational singly-excited initial state, $\Psi^{(r)}_0$, with vertical time-axis and horizontal $\theta$-axis on upper, $\varepsilon_{U_1}(\theta,\phi)$ (top row), middle, $\varepsilon_{M_1}(\theta,\phi)$ (middle row) and lower, $\varepsilon_{L_1}(\theta,\phi)$ (bottom row),  vibro-polaritonic potential energy surfaces for $\eta=0.05$ (left column) and $\eta=0.1$ (right column). Intersection at $\theta=\frac{\pi}{2}$ indicated by blue vertical line. Total density, $\rho^{(r)}(\theta,t)=\sum_l \rho^{(r)}_l(\theta,t)$, with $l=G,L_1,M_1,U_1$ normalized for fixed time $t$.
}
\label{fig.rotential_density_dynamics_molXite}
\end{figure}

In Fig.\ref{fig.rotential_density_dynamics_molXite}, the time-evolution of $\rho^{(r)}_l(\theta,t)$ is shown for excited-state surfaces $\varepsilon_{U_1}$ (top row), $\varepsilon_{M_1}$ (middle row) and $\varepsilon_{L_1}$ (bottom row), respectively, with $\eta=0.05$ (left column) and $\eta=0.1$ (right column). The horizontal $\theta$-axis runs from $\pi$ to $0$ (left to right) and the vertical time-axis runs from $0\,\mathrm{fs}$ to $3000\,\mathrm{fs}$ (top to bottom). The position of the intersection point at $\theta=\frac{\pi}{2}$ is marked by a vertical blue line. 
At $t=0$, both $L_1$- and $U_1$-surfaces are nearly equivalently populated (in line with Figs.\ref{fig.rovib_polariton_dynamics}(a) and (b)), and $\rho^{(r)}_{L_1}(\theta,t)$ and $\rho^{(r)}_{U_1}(\theta,t)$ show two symmetric maxima along $\theta$ with respect to the intersection point. 

In the VSC regime with $\eta=0.05$ (\textit{cf.} Fig.\ref{fig.rotential_density_dynamics_molXite}, left column), density on the $U_1$-surface symmetrically approaches the VPCI and is transferred to the $L_1$-surface with minor contributions on $M_1$. In parallel, $\rho^{(r)}_{L_1}(\theta,t)$ tends away from the intersection point and develops a rich interference pattern as time evolves. This pattern can be rationalized in terms of rotational wave packet interference, where non-adiabatic transfer between $U_1$- and $L_1$-surfaces induces interference between wave packets initially located on different surfaces. At later times, several recurrences at the $U_1$-surface are observed due to $\rho^{(r)}_{L_1}(\theta,t)$ reentering the intersection region. Notably, significant density on $\varepsilon_{M_1}$ is mainly found in the harmonic region close to the VPCI. For VUSC with $\eta=0.1$, the density transfer between $L_1$- and $U_1$-surfaces is enhanced but qualitatively equivalent to the VSC regime.
\begin{figure}[hbt]
\includegraphics[scale=1.0]{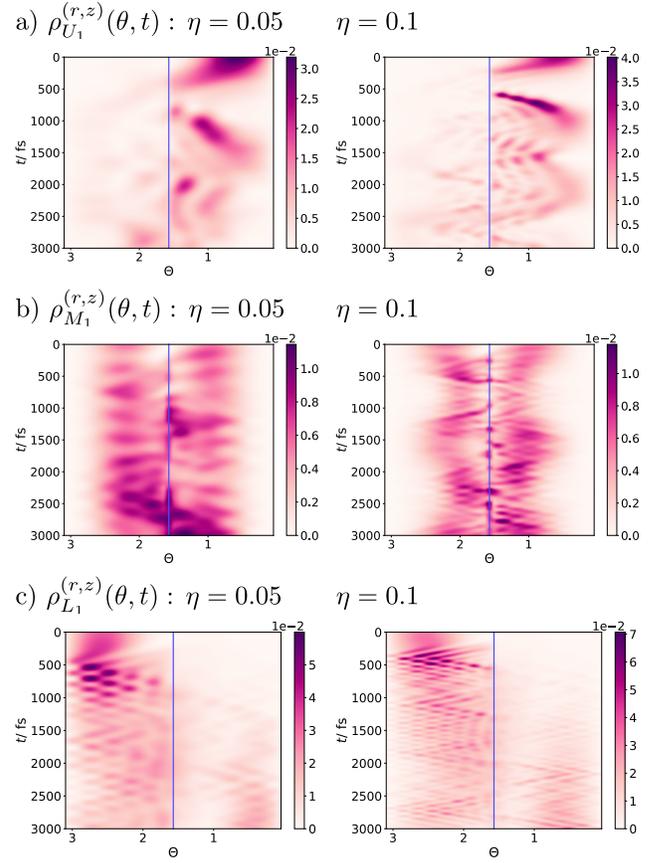}
\renewcommand{\baselinestretch}{1.}
\caption{Time-evolution of adiabatic reduced rotational densities, $\rho^{(r,z)}_l(\theta,t)$, for singly-excited superposition state, $\Psi^{(r,z)}_0$, with vertical time-axis and horizontal $\theta$-axis on upper, $\varepsilon_{U_1}(\theta,\phi)$ (top row), middle, $\varepsilon_{M_1}(\theta,\phi)$ (middle row) and lower, $\varepsilon_{L_1}(\theta,\phi)$ (bottom row),  vibro-polaritonic potential energy surfaces for $\eta=0.05$ (left column) and $\eta=0.1$ (right column). Intersection at $\theta=\frac{\pi}{2}$ indicated by blue vertical line. Total density, $\rho^{(r,z)}(\theta,t)=\sum_l \rho^{(r,z)}_l(\theta,t)$, with $l=G,L_1,M_1,U_1$ normalized for fixed time $t$.
}
\label{fig.rotential_density_dynamics_sup1Xite}
\end{figure}

Turning to the initial superposition state, $\Psi^{(r,z)}_0$, we find the reduced vibro-polaritonic rotational density to be initially distributed over all three vibro-polaritonic surfaces (\textit{cf.} Fig.\ref{fig.rotential_density_dynamics_sup1Xite}) in line with the population analysis. Further at $t=0$, densities $\rho^{(r,z)}_{L_1}(\theta,t)$ and $\rho^{(r,z)}_{U_1}(\theta,t)$ show a highly asymmetric distribution with respect to the VPCI, in contrast to the symmetric character of $\rho^{(r,z)}_{M_1}(\theta,t)$.

As time-evolves, we again observe a non-adiabatic density transfer via the VPCI between $U_1$- and $L_1$-surfaces for both interaction regimes (\textit{cf.} Fig.\ref{fig.rotential_density_dynamics_sup1Xite}, left and right column). The transfer is enhanced for VUSC in line with the population analysis given above and characterized by a rich interference pattern. Further, the asymmetric character of the reduced densities on the $U_1$- and $L_1$-surfaces is clearly observable for the whole time-interval studied and the $M_1$-surface is substantially stronger explored for both coupling regimes. We attribute the latter effect to a finite initial population of $\ket{M_1}$, as the non-adiabatic population transfer to $\varepsilon_{M_1}$ is rather inefficient as seen before.

In summary, rotational densities allow to reveal the non-adiabatic character of vibro-polaritonic population transfer, which is dominated by (i) a funneling effect of the VPCI and (ii) interference of rotational wave packets initially located on different vibro-polaritonic surfaces.

\subsection{Rovibro-Polaritonic Infrared Spectra}
\label{subsec.rovib_spectra}
We now consider the spectroscopic characterization of the rovibro-polaritonic system with an emphasis on non-adiabatic signatures of the three-state VPCI. Infrared (IR) spectra, $\sigma(\omega)$, are calculated as
\begin{equation}
\sigma(\omega)
= 
A 
\displaystyle\int^\infty_0
C(t)\,e^{\mathrm{i}\omega\,t}
\mathrm{d}t
\quad,
\label{spec}
\end{equation}
with an autocorrelation function, $C(t)$, and a constant here chosen as$A=1$. The autocorrelation function $C(t)$ is defined as the overlap between the initial states $\Psi(0)=\Psi_0^{(r)}$ and $\Psi_0^{(r,z)}$ as defined in Eqs.(\ref{eq.init_molxite}) and (\ref{eq.init_supxite}) and the propagated state, $\Psi(t)$, under the influence of the coupled Pauli-Fierz Hamiltonian. (Note that  this is an approximation/variant to the ``usual'' procedure in which an eigenstate of the Hamiltonian is multiplied by the dipole function and then propagated in time.\cite{heller}) We obtain two different spectra due to the two initial states, or implicitly, molecule-cavity dipole functions $\underline{\mu}$ (see above). Further details on the evaluation of $C(t)$ and $\sigma(\omega)$ can be found in Appendix A.
\begin{figure*}[hbt]
\includegraphics[scale=1.0]{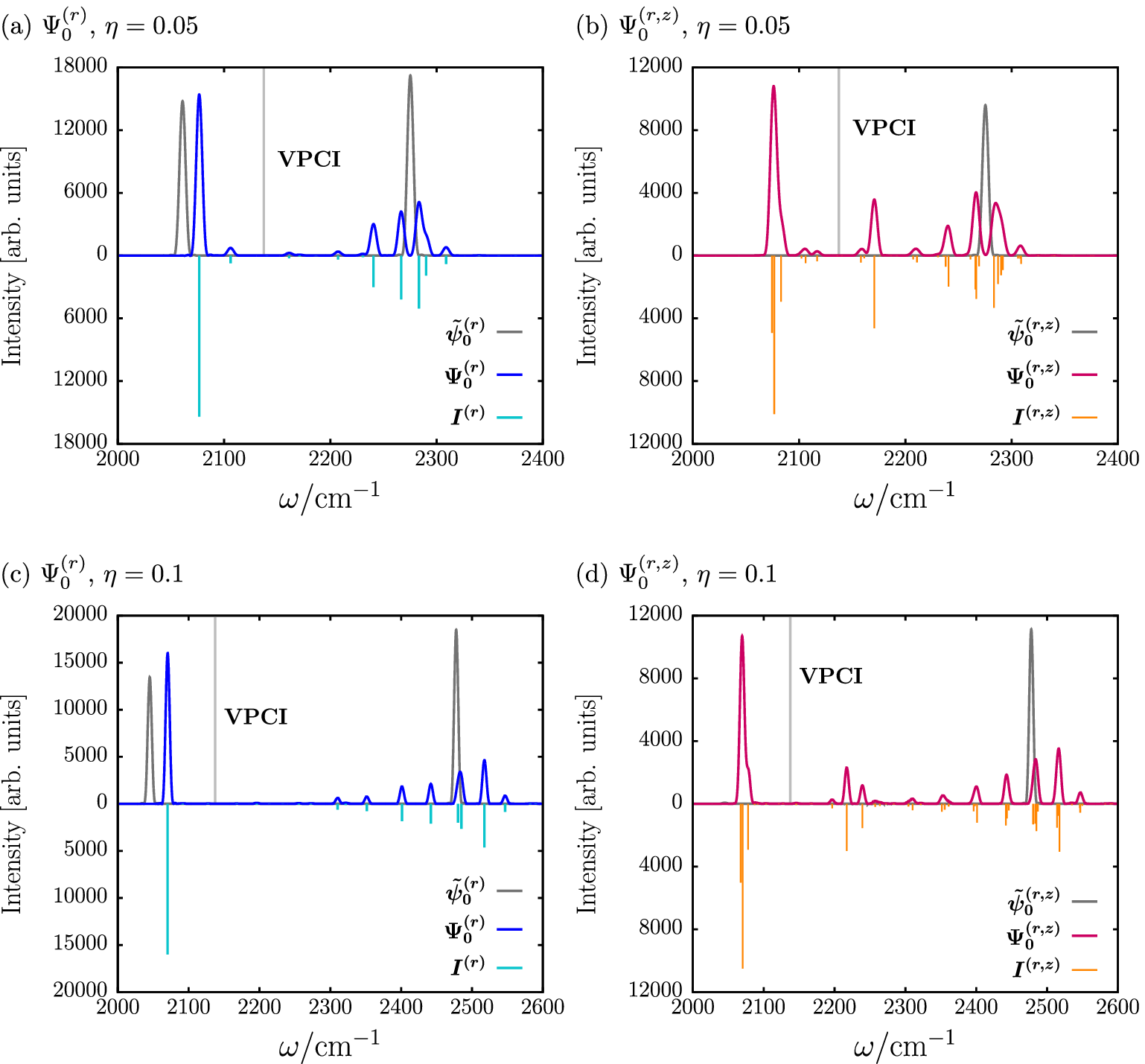}
\renewcommand{\baselinestretch}{1.}
\caption{Vibro-polaritonic infrared spectra for different initial states and light-matter interaction parameters $\eta$ with single-mode limit initial states, $\tilde{\psi}^{(r)}_0=\ket{1_r,0_z}$, and, $\tilde{\psi}^{(r,z)}_0=\frac{1}{\sqrt{2}}\left(\ket{1_r,0_z}+\ket{0_r,1_z}\right)$, (grey curves), VPCI energy (grey vertical lines) and intensities, $I^{(r)}\propto\vert\braket{\Psi^{(r)}_0\vert\phi_i}\vert^2$, and, $I^{(r,z)}\propto\vert\braket{\Psi^{(r,z)}_0\vert\phi_i}\vert^2$. Top row: Infrared-spectra for the vibrational strong coupling (VSC) regime with $\eta=0.05$ for (a) molecular rovibrationally excited initial state, $\Psi^{(r)}_0$, and (b) light-matter superposition state, $\Psi^{(r,z)}_0$. Bottom row: (c) and (d) analogous to (a) and (b) in top row for onset of vibrational ultrastrong strong coupling (VUSC) regime at $\eta=0.1$.}
\label{fig.lici_spec_fig}
\end{figure*}

All spectra shown below have been obtained for a total propagation time, $t_f=6000\,\mathrm{fs}$. In order to reveal rotational effects and effects of a second cavity mode, we consider IR spectra obtained for a purely vibro-polaritonic system in the single-cavity-mode limit as a reference, \textit{i.e.}, a CO molecule solely interacting with the $z$-polarized cavity-mode. Here, we take into account initial states $\Psi(0)=\tilde{\psi}^{(r)}_0=\ket{1_r,0_z}$ and $\Psi(0)=\tilde{\psi}^{(r,z)}_0=\frac{1}{\sqrt{2}}\left(\ket{1_r,0_z}+\ket{0_r,1_z}\right)$. In order to interpret spectroscopic signatures, we also consider contributions of eigenstates, $\phi_i$, of the effective Hamiltonian $\underline{\underline{H}}$ ({\it{cf.}} Eq.(\ref{eq.matrix_rovib_pauli_fierz_hamiltonian})), 
to initial states, as given by intensities $I^{(r)}\propto\vert\braket{\Psi^{(r)}_0\vert\phi_i}\vert^2$ and $I^{(r,z)}\propto\vert\braket{\Psi^{(r,z)}_0\vert\phi_i}\vert^2$, respectively (details are provided in Appendix A).
\subsubsection{Infrared Spectra}
\label{subsubsec.rovib_ir_spectra}
We first consider spectra in the VSC regime with $\eta=0.05$ and initial states $\Psi^{(r)}_0$ and $\Psi^{(r,z)}_0$ as shown in Fig.\ref{fig.lici_spec_fig}(a) and (b). For both initial states, we observe a series of transitions with a dominant $L_1$-peak at $2076\,\mathrm{cm}^{-1}$ below the intersection point energy of $2137\,\mathrm{cm}^{-1}$. In the region of the $U_1$-surface, we find a progression of five peaks between $2207\,\mathrm{cm}^{-1}$ and $2309\,\mathrm{cm}^{-1}$ depending on the initial state with spacing $17-33\,\mathrm{cm}^{-1}$, which decreases for increasing peak energy. The peak intensity increases with energy with a significantly less intense high-energy peak terminating the progression. Notably, the latter transitions do not reflect purely rovibrational states, but as discussed below correspond to rovibro-polaritonic hybrid states. Moreover, for $\Psi^{(r,z)}_0$ a prominent peak is found at $2171\,\mathrm{cm}^{-1}$, which resembles an excitation of the purely photonic intermediate polariton state and is absent for $\Psi^{(r)}_0$. 

Further, the single-cavity-mode limit reveals the well-known $L_1$- and $U_1$-peaks for $\tilde{\psi}^{(r)}_0$ with Rabi-splitting $\Omega_R=214\,\mathrm{cm}^{-1}$, respectively (grey curves). For $\tilde{\psi}^{(r,z)}_0$, the chosen linear combination exactly captures the $U_1$-state such that only a single peak is observed here. Accordingly, the inclusion of rotational effects leads to a much richer excitation spectrum (colored curves) than accessible in the single-mode limit (grey curves). Explicitly, non-adiabatic effects show up as multi-peak progression in the spectral region of the upper vibro-polaritonic state, which is characterized by the inverse cone topology of the $\varepsilon_{U_1}$-surface.  

From an analysis of intensities, $I^{(r)}$ and $I^{(r,z)}$, we are able to reveal the detailed character of observed excitations, which are hidden in spectra due to finite peak widths resulting from finite propagation times of rovibro-polaritonic wave functions. Both, $I^{(r)}$ and $I^{(r,z)}$ exhibit similar dominant contributions to the light-matter hybrid $L_1$- and $U_1$-peaks, whereas only $I^{(r,z)}$ exhibits the purely photonic $M_1$-peak as expected. Notably, $I^{(r,z)}$ shows additional contributions from several energetically close lying states with lower intensity to individual spectroscopically resolved peaks. We attribute a dominant rotational light-matter hybrid character to the latter as they are absent for the rovibrationally excited initial state, \textit{i.e.}, seem to be dominated by a photonic contribution. 

Turning to IR spectra at the onset of the VUSC regime with $\eta=0.1$, we observe a broadening of the spectrum and additional peaks emerge as shown in Fig.\ref{fig.lici_spec_fig}(c) and (d). In the single-mode limit, we find a large Rabi-splitting of $\Omega_R=433\,\mathrm{cm}^{-1}$. Further, the progression in the $U_1$-spectral region related to rovibro-polaritonic states is now more pronounced with seven peaks showing slightly increased spacing of $29-49\,\mathrm{cm}^{-1}$, which again decreases for increasing peak energy. The middle-polariton peak at $2171\,\mathrm{cm}^{-1}$ for the light-matter excited initials state splits into three peaks at $2116,2239$ and $2258\,\mathrm{cm}^{-1}$, respectively. As previously for VSC, we observe a number of (potentially energetically close) rovibro-polaritonic hybrid states with lower intensity, which contribute to peaks in the spectrum obtained from $\Psi^{(r,z)}_0$ as indicated by $I^{(r,z)}$ and $I^{(r)}$. 

In conclusion, IR spectra reveal the formation of light-matter hybrid states with contributions from both molecular rotations and vibrations as well as cavity mode excitations, \textit{i.e.}, rovibro-polaritonic states, with non-adiabatic signatures prominently manifesting as rovibro-polaritonic progression in the spectral region of the $U_1$-inverse cone. 
\subsubsection{Cavity Loss Effects}
\label{subsubsec.lossy_cavity_rovib_spectra}
Finally, we consider the impact of experimentally ubiquitous spontaneous emission effects from cavity modes on rovibro-polaritonic infrared spectra. Spontaneous emission manifests in finite excited cavity mode lifetimes (Finite lifetimes of molecular rovibrational states are neglected in what follows). These are considered here by a phenomenological approach recently employed in electronic strong coupling studies\cite{ulusoy2020,felicetti2020,antoniou2020}. There, the cavity mode Hamiltonian, $\hat{H}_C$, is replaced by a non-Hermitian operator
\begin{equation}
\hat{H}^{(\kappa)}_C
=
\sum_{\lambda=z,y}
\left(
\hbar\omega_c
-
\mathrm{i}\,
\frac{\kappa}{2}
\right)
\hat{a}^\dagger_\lambda
\hat{a}_\lambda
\quad .
\label{eq.nonhermitian_cavity_hamiltonian}
\end{equation}
The imaginary contribution accounts for an effective cavity decay rate, $\kappa$, which we here set to  $\kappa=43\,\mathrm{cm}^{-1}$ to model an infrared cavity with quality factor $Q=\frac{\hbar\omega_c}{\kappa}=50$, being slightly lower as in Ref.\cite{shalabney2015}. Note that the zero-point energy contribution is neglected in Eq.\ref{eq.nonhermitian_cavity_hamiltonian} to avoid artificial ground state decay\cite{ulusoy2020}.  
\begin{figure*}[hbt]
\includegraphics[scale=1.0]{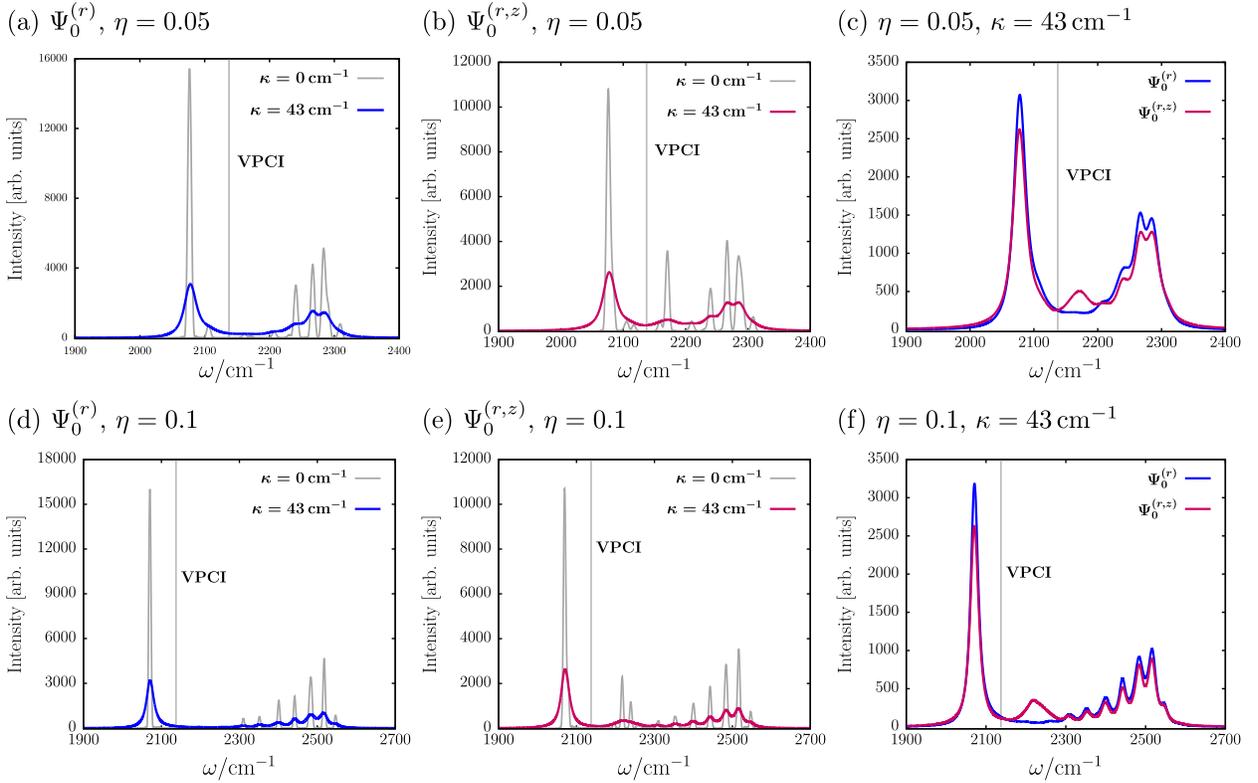}
\renewcommand{\baselinestretch}{1.}
\caption{Vibro-polaritonic infrared spectra subject to dissipation (colored) with cavity decay rate, $\kappa$, for different initial states and light-matter interaction regimes with non-dissipative reference (grey) and VPCI energy (vertical line). Top row: Infrared-spectra for the vibrational strong coupling (VSC) regime with $\eta=0.05$ for (a) molecular rovibrationally excited initial state $\Psi^{(r)}_0$, (b) light-matter superposition state $\Psi^{(r,z)}_0$ and (c) comparison of both dissipative spectra. Bottom row: (d)-(f) analogous to (a)-(c) in top row for onset of vibrational ultrastrong strong coupling (VUSC) regime at $\eta=0.1$.}
\label{fig.lici_spec_diss_fig}
\end{figure*}

In Fig.\ref{fig.lici_spec_diss_fig}, we compare infrared spectra subject to spontaneous emission (dissipation) with non-dissipative results as presented in Fig.\ref{fig.lici_spec_fig}. Most prominently, cavity-loss effects manifest themselves in a significant intensity reduction and peak broadening, especially for the $M_1$-peak themselves . The latter again supports the purely photonic nature of the $\ket{M_1}$-state. For the coupling scenario $\eta=0.05$, spectra resulting from $\Psi^{(r)}_0$ and $\Psi^{(r,z)}_0$ differ only slightly in their intensity as depicted in Figs.\ref{fig.lici_spec_diss_fig}(a)--(c), with slightly more intense $L_1$-/$U_1$-transitions for $\Psi^{(r)}_0$. Further, the rovibro-polaritonic progression induced by the VPCI is only weakly observable and resembles several shoulders in the $U_1$-spectra region. For increased light-matter interaction strength (\textit{cf.} Figs.\ref{fig.lici_spec_diss_fig}(d)--(f)), the progression is clearly visible for both initial state although it suffers from the peak broadening effects.  
\section{Summary and Outlook}
\label{sec.summary}
We studied a model system composed of a single rovibrating diatomic molecule (carbon monoxide), which interacts with two energetically degenerate and orthogonally polarized optical modes of a Fabry-P\'erot-type cavity. The cavity modes were tuned resonant to the fundamental vibrational transition of the CO stretching mode and we discussed the vibrational strong coupling regime up to the onset of the vibrational ultrastrong coupling regime. Based on an energetically motivated adiabatic separation of rotational and vibro-polaritonic degrees of freedom, we identified a three dimensional single-excitation manifold of vibro-polaritonic surfaces, which is subject to two distinct three-state vibro-polaritonic conical intersections in a two-dimensional angular coordinate branching space. The lower and the upper vibro-polaritonic states are of mixed light-matter hybrid character and the middle polaritonic state is purely photonic in nature. In the VSC regime, the lower and upper vibro-polaritonic surfaces exhibit a symmetric double-cone topology with respect to a harmonic middle polaritonic surface at the triply degenerate intersection point. For increasing light-matter interaction, the symmetric splitting of upper and lower vibro-polaritonic surfaces becomes strongly asymmetric, with the inverse cone of the upper vibro-polaritonic surface dominating the VPCI topology and a mild ``Mexican-hat''-type topology in the lower vibro-polaritonic surface.

From a dynamical perspective, we studied the time-evolution of vibro-polaritonic populations and reduced vibro-polaritonic rotational densities. The population dynamics reveal, that the presence of rotational degrees of freedom induces an efficient transfer channel between vibro-polaritonic states in terms of the VPCI. Moreover, the rotational density shows a rich interference pattern throughout the dynamics, due to interfering rotational wave packets initially located on different vibro-polaritonic surfaces.

Further, we studied infrared spectra of the rovibrating light-matter hybrid system with respect to different initial conditions, which mimicked classical laser-field excitations of a (i) molecular rovibrational excited state and (ii) a light-matter superposition state. We observe bright transitions to all three vibro-polaritonic states, only when the laser field is allowed to initially excite a light-matter superposition state. The non-trivial topology of the VPCI manifest as a pronounced progression in the spectral region of the upper vibro-polaritonic surface, which we find to correspond to rovibro-polaritonic states containing contributions of molecular rotational, vibrational and cavity degrees of freedom. Further, we considered finite lifetimes of cavity mode excitations due to spontaneous emission effects, which manifest in a significant reduction of the purely photonic middle polariton peak and broadening of the rovibro-polaritonic progression. 

Possible extensions of the present study refer to increasing the reduced diabatic ``vibro-polaritonic'' basis to go beyond the (effective) single-molecule picture and aim at the study of collective effects. In particular, it would be instructive to investigate how the orientation dependent light-matter interaction effects the nature of ``dark states'' in rovibro-polaritonic problems. Further, we did not consider the nature of the vibro-polaritonic ground state here, which renders central for purely rotational dynamics of molecules in infrared Fabry-P\'erot cavities. Moreover, we note that for electronic strong coupling (ESC) problems, it had already been suggested in Refs.\cite{triana2018,triana2019} that (single-mode) cavity induced CIs could be formed, how one could detect them and what is the influence of molecular rotation on them\cite{triana2019}. This work is complemented by Ref.\cite{farag2021}, where a similar problem has been studied with respect to Berry Phase effects in a polaritonic CI setting. Here, differences to ``classical'' light-induced conical intersections\cite{sindelka2011}, which emerge from a linearly polarized, classical laser field with a \textit{single} polarization direction, could be expected in the fully quantized cQED setting of ESC scenarios featuring (at least) \textit{two} degenerate, orthogonally polarized cavity modes. Finally, the study of rovibro-polaritonic effects on cavity-altered chemical reactions and on polaritonic infrared spectra might be rewarding for a deeper understanding of molecular cQED.

\section*{Acknowledgements}
We acknowledge fruitful discussions with Shreya Sinha, Foudhil Bouakline, David Picconi and Tillmann Klamroth (all from Potsdam). The authors thank the Deutsche Forschungsgemeinschaft (DFG) for financial support through project Sa 547/18. E.W. Fischer also acknowledges support by the International Max Planck Research School for Elementary Processes in Physical Chemistry.

\section*{Data Availability Statement}
The data that support the findings of this study are available from the corresponding author upon reasonable request.

\section*{Conflict of Interest}
The authors have no conflicts to disclose.

\section*{Appendix}
\subsection{Numerical Details}
\setcounter{equation}{0}
\renewcommand{\theequation}{\thesubsection\arabic{equation}}
We employ the MCTDH approach\cite{meyer2012,meyer1990,manthe1992,beck2000,meyer2003} as implemented in the Heidelberg MCTDH package, version 8.6.0\cite{mctdh2019}. The state-specific rotational wave packets $\varphi_k(\theta,\phi,t)$ in Eq.\eqref{eq.rovibpol_wavepacket} are here expanded as
\begin{equation}
\varphi_k(\theta,\phi,t)
=
\sum^{n}_{i=1}
A_{k i}(t)\,
\varphi^{(k)}_i(\theta,\phi,t)
\quad,
\end{equation}
with time-dependent coefficients, $A_{ki}(t)$, and state- and time-dependent rotational single-particle functions (SPFs), $\varphi^{(k)}_i(\theta,\phi,t)$, respectively. For all calculations, we chose $n=3$ SPFs for each diabatic state and represented the states-specific rotational wave packets in a two-dimensional Legendre discrete variable representation (PLeg) with $N_\theta=51$ and $N_\phi=37$ grid points. 

Vibro-polaritonic (adiabatic) populations $P^{(l)}_\mathrm{ad}(t)$ are obtained from Eq.\eqref{eq.diabatic_populations} by taking into account the adiabatic representation of the zero-order rotational wave packets, $\varphi_k(\theta,\phi,t)$, given as
\begin{align}
\varphi^\mathrm{ad}_l(\theta,\phi,t)
&=
\sum_k
u_{lk}(\theta,\phi)\,
\varphi_k(\theta,\phi,t)
\quad,
\label{eq.adiabatic_rotational_states}
\end{align}
where $u_{lk}(\theta,\phi)$ are elements of a unitary matrix $\underline{\underline{U}}(\theta,\phi)$, which diagonalizes the potential in Eq.\eqref{eq.potential_matrix}. The vibro-polaritonic (adiabatic) rotational wave packets, $\varphi^\mathrm{ad}_l(\theta,\phi,t)$, evolve on the surfaces as depicted in Fig.\ref{fig.lici_free_vsc_fig}. Further, infrared spectra are calculated as (setting a possible prefactor to 1)
\begin{equation}
\sigma(\omega)
=
\displaystyle\int^T_0
C(t)\,
f_W(t)\,
e^{\mathrm{i}\omega\,t}
\mathrm{d}t \quad ,
\end{equation}
with autocorrelation function $C(t)$ and window function 
\begin{equation}
f_W(t)
=
\left(1-\dfrac{t}{T}\right)
\cos\left(\dfrac{\pi}{T}\,t\right)
+
\dfrac{1}{\pi}
\sin\left(\frac{\pi}{T}\,t\right) \quad .
\end{equation}
For the Hermitian, non-dissipative case, we propagate a given initial wave packet $\Psi(0)$ under the influence of the Pauli-Fierz Hamiltonian in Eq.\eqref{eq.rovib_pauli_fierz}, to a time $t_f=3000\,\mathrm{fs}$ and the spectrum is obtained with $T=2\,t_f=6000\,\mathrm{fs}$ since the autocorrelation function is evaluated as\cite{beck2000}
\begin{equation}
C(t)
=
\braket{\Psi^\star(t/2)\vert \Psi(t/2)}
\quad.
\end{equation} 
In the non-Hermitian scenario, $\hat{H}^{(\kappa)}_C$ as defined in Eq.\eqref{eq.nonhermitian_cavity_hamiltonian} is used instead for the cavity modes, and the autocorrelation function is computed as
\begin{equation}
C(t)
=
\braket{\Psi(0)\vert \Psi(t)}
\quad,
\end{equation} 
where we propagate up to $T=t_f=6000\,\mathrm{fs}$. Finally, for analysis, intensities (stick spectra)
 are obtained as
\begin{equation}
\begin{matrix}
I^{(r)}
=
\sum_i
\vert
\braket{\Psi^{(r)}_0
\vert
\phi_i}
\vert^2\,
\delta(\hbar\omega-\varepsilon_i)
\vspace{0.2cm}
\\
I^{(r,z)}
=
\sum_i
\vert
\braket{\Psi^{(r,z)}_0
\vert
\phi_i}
\vert^2\,
\delta(\hbar\omega-\varepsilon_i)
\end{matrix}
\quad,
\end{equation}
with eigenvalues $\varepsilon_i$ and eigenstates $\phi_i$ of $\underline{\underline{H}}$ calculated by means of a Lanczos algorithm as implemented in the Heidelberg MCTDH package. All converged results have been obtained based on a two-dimensional Legendre discrete variable representation (PLeg) with $N_\theta=51$ and $N_\phi=37$ grid points as above, with $9000$ iteration steps.

\subsection{3-State-VPCI Details}
\setcounter{equation}{0}
\renewcommand{\theequation}{\thesubsection\arabic{equation}}
We expand trigonometric functions in $\underline{\underline{V}}(\theta,\phi)$ up to first order in $\theta$ and $\phi$ around the intersection point at $(\theta,\phi)=(\frac{\pi}{2},0)$,  as
\begin{equation}
\cos\theta
\approx
\theta-\dfrac{\pi}{2}
\quad,
\hspace{0.5cm}
\sin\theta\sin\phi
\approx
\phi
\quad,
\end{equation}
and
\begin{equation}
f(\theta,\phi)
\approx
\phi^2
+
\left(
\theta-\dfrac{\pi}{2}
\right)^2
=
f^{(1)}(\theta,\phi)
\quad.
\end{equation}

The resulting approximate potential energy matrix is given by
\begin{widetext}
\begin{equation}
\underline{\underline{V}}^{(1)}(\theta,\phi)
=
\begin{pmatrix}
G_{00}\,f^{(1)}(\theta,\phi) & G_{10}\,f^{(1)}(\theta,\phi) & -g_{00}\,\left(\theta-\dfrac{\pi}{2}\right) & g_{00}\,\phi
\vspace{0.2cm}
\\
G_{10}\,f^{(1)}(\theta,\phi) & \hbar\omega_{10}+G_{11}\,f^{(1)}(\theta,\phi) & -g_{10}\,\left(\theta-\dfrac{\pi}{2}\right) & g_{10}\,\phi
\vspace{0.2cm}
\\
-g_{00}\,\left(\theta-\dfrac{\pi}{2}\right) &
-
g_{10}\,\left(\theta-\dfrac{\pi}{2}\right) & \hbar\omega_c+G_{00}\,f^{(1)}(\theta,\phi) & 0
\vspace{0.2cm}
\\
g_{00}\,\phi & g_{10}\,\phi & 0 & \hbar\omega_c+G_{00}\,f^{(1)}(\theta,\phi)
\end{pmatrix}
\quad.
\label{eq.linear_coupling_lici_potential}
\end{equation}
\end{widetext}
with $\hbar\omega_{10}=\hbar\omega_c$, $G_{ij}=\frac{g^2}{\hbar\omega_c}\braket{d^2}_{ij}$ as well as $g_{ij}=g\,d_{ij}$ where $i,j=0,1$. The linear approximation, holds for small deviations from the intersection coordinate where $f^{(1)}(\theta,\phi)\approx0$. For larger values of $\theta$ and $\phi$, the quadratic approximation then takes into account the DSE-related term, $f^{(1)}(\theta,\phi)$. The corresponding eigenvalues of $\underline{\underline{V}}^{(1)}(\theta,\phi)$ correspond to the different approximations of vibro-polaritonic surfaces with cuts shown in Fig.\ref{fig.lici_free_vsc_fig}, bottom-row.


\begin{thebibliography}{100}
\bibitem{rovibXite} We assume, $\mu^{(r)}=d^{(r)}_{10}\ket{1_r,0_z,0_y}\bra{0_r,0_z,0_y}\cos\theta$, with, $\mu^{(r)}\ket{0_r,0_z,0_y}Y^0_0=\Psi^{(r)}_0$, and, $\mu^{(r,z)}=d^{(r,z)}_{10}\left(\ket{1_r,0_z,0_y}\bra{0_r,0_z,0_y}+\ket{0_r,1_z,0_y}\bra{0_r,0_z,0_y}\right)\cos\theta$, such that, $\mu^{(r,z)}\ket{0_r,0_z,0_y}Y^0_0=\Psi^{(r,z)}_0$. For both, we set, $d^{(r)}_{10}=d^{(r,z)}_{10}=1$.
\end{thebibliography}
\end{document}